\documentclass[prd,aps,preprint,tightenlines,superscriptaddress]{revtex4}
\usepackage{epsfig}
\preprint{DOE/ER/40762-298} \preprint{UM-PP\#04-007}

\def \to {\rightarrow}

\def \pn3 {\psi_{u\bar d g}}

\begin{document}
\title{Generalized Parton Distributions at $x\to 1$}
\author{Feng Yuan} \email{fyuan@physics.umd.edu} \affiliation{Department of Physics,
University of Maryland, College Park, Maryland 20742}
\date{\today}
\vspace{0.5in}
\begin{abstract}

Generalized parton distributions at large $x$ are studied in
perturbative QCD approach. As $x\to 1$ and at finite $t$, there is
no $t$ dependence for the GPDs which means that the active quark
is at the center of the transverse space. We also obtain the power
behavior: $H_q^\pi(x,\xi,t)\sim (1-x)^2/(1-\xi^2)$ for pion;
$H_q(x,\xi,t)\sim (1-x)^3/(1-\xi^2)^2$ and $E_q(x,\xi,t)\sim
(1-x)^5/(1-\xi^2)^3f(\xi)$ for nucleon, where $f(\xi)$ represents
the additional dependence on $\xi$.
\end{abstract}

\maketitle

In recent years, there has been considerable interest in
generalized parton distributions (GPDs)
\cite{{Ji:1996ek},{Muller:1998fv},{Radyushkin:1997ki}}, which were
introduced originally to understand the quark and gluon
contributions to the proton spin \cite{Ji:1996ek}. They are also
related to the quantum phase space distributions of partons in the
hadrons \cite{wigner}. The theoretical framework of the GPDs and
their implications about the deeply virtual Compton scattering,
deeply virtual meson production, and the doubly-virtual Compton
scattering have been well established
\cite{{Ji:1998pc},{Radyushkin:2000uy},{Goeke:2001tz},{Belitsky:2001ns},
{Diehl:2003ny},{Guidal:2002kt},{Belitsky:2002tf}}. Apart from the
renormalization scale, the GPDs depend on the momentum transfer
$t$, the light-cone momentum fraction $x$, and the skewness
parameter $\xi$ which measures the momentum transfer along the
light-cone direction. In phenomenology, the GPDs are parameterized
through the double-distributions \cite{Radyushkin:1998bz} and fit
to the experimental data
\cite{{Goeke:2001tz},{Belitsky:2001ns},{Diehl:2003ny},
{Guidal:2002kt},{Belitsky:2002tf}}. However, these
parameterizations have too much freedom, and we still have a long
way to go for a complete understanding of the GPDs. In this
context, any theoretical result on the behavior of GPDs will
provide important information. For example, the polynomality
condition \cite{{Ji:1998pc}}, and the positivity constraints
\cite{Pobylitsa:2001nt} have already played significant roles in
the parametrizations of GPDs. The light-cone framework provides
useful guidelines for calculating the GPDs once the wave functions
are known \cite{Brodsky:2000xy}. More recently, the GPDs at large
$t$ have been explored \cite{Hoodbhoy:2003uu}, yielding important
constraints as well.

In this paper, we study the GPDs in the kinematic limit of $x\to
1$. For the forward parton distribution, a power behavior at large
$x$ was predicted based on the power counting rules, for example,
$(1-x)^2$ for pion, and $(1-x)^3$ for nucleon
\cite{{Drell:1969km},{West:1970av},Brodsky:1973kr,Matveev:ra,Farrar:yb,{Brodsky:1994kg}}.
This power behavior comes from the fact that the hard gluon
exchanges dominate the structure functions at $x\to 1$, and is
calculable in perturbative QCD \cite{Lepage:1980fj,Mueller:sg}. In
this paper, we will follow these ideas to analyze the dependence
of GPDs on the three variables $x$, $\xi$ and $t$ in the limit of
$x\to 1$. We use the QCD factorization approach, and express the
GPDs in terms of the distribution amplitudes of hadrons. In the
limit of $x\to 1$, the power behavior of the GPDs does not depend
on a particular input of the distribution amplitudes, and
therefore can be predicted model-independently. More importantly,
the $\xi$ and $t$ dependences can also be calculated. For example,
we find that there is no $t$-dependence at $x\to 1$, which agrees
with the previous intuitions \cite{Burkardt:2003ck}.

We take $(1-x)$ as a small parameter, and expand the GPDs in terms
of $(1-x)$. In the process, we assume the variables $\xi$ and $t$
finite. Finite $\xi$ means $\xi<x$ and restricts our analysis
valid in the DGLAP region for the GPDs. The relevant Feynman
diagrams are shown in Fig.~1 for pion, and in Fig.~2 for nucleon
for a typical contribution. The variables $P$ and $P'$ are the
initial and final state hadron momenta, respectively, and
$t=\Delta^2=(P-P')^2$. We further introduce two vectors $\overline
P$ and $n$: $\overline{P}=(P+P')/2$, $n^2=0$, and $n\cdot
\overline{P}=1$. The skewness parameter is defined as $\xi
=-n\cdot (P'-P)/2$. The initial and final light-cone momenta of
the quarks are then $(x+\xi)$ and $(x-\xi)$, respectively. In the
following, we will neglect the masses of the hadrons, and then
$t=-\vec \Delta_\perp^2/(1-\xi^2)$ where $\vec\Delta_\perp$ is the
transverse part of the momentum transfer $\Delta$.

As shown in Fig.~1, the intermediate state has momentum $k$ which
will be integrated out. To avoid an infrared divergence, we keep
$k_\perp$ much larger than $\Lambda_{\rm QCD}$. The offshellness
of the quark and gluon propagators are on the order of $\vec
k_\perp^2/(1-x)$. So, we have the following hierarchy of scales in
the limit of $x\to 1$: $\vec k_\perp^2/(1-x)\gg \vec
k_\perp^2\gg\Lambda_{\rm QCD}^2$, and $\vec k_\perp^2/(1-x)\gg
(-t)$ as well. These relations will be used to get the
leading-order results, and any higher power in $(1-x)$ will be
neglected.

The GPD $H(x,\xi,t)$ for pion is defined as
\begin{eqnarray}
H_q(x,\xi,t)&&\!\!=\!\!\frac{1}{2}\int \frac{d\lambda}{2\pi}
e^{i\lambda x} \left\langle
\pi;P'\left|\overline{\psi}_q\left(-\frac{\lambda}{2}n\right)\!\!
\not\! n  {\cal L}
\psi_q\left(\frac{\lambda}{2}n\right)\right|\pi;P\right\rangle
\nonumber\ , \label{pion}
\end{eqnarray}
where ${\cal L}$ represents the light-cone gauge link. We work in
Feynman gauge, and the leading-order diagrams were shown in
Fig.~1, where the double lines represent the eikonal contributions
from the gauge link, and the cross indicates the intermediate
state on mass shell. The initial and final states are replaced by
the light-cone Fock component of hadrons with the minimal number
of partons. After integrating over the internal transverse
momentum $l_\perp$, the light-cone wave function leads to the
distribution amplitude,
$\phi(x)=\int\frac{d^2l_\perp}{(2\pi)^3}\psi(x,l_\perp)$.

\begin{figure}[t]
\begin{center}
\includegraphics[height=5.0cm]{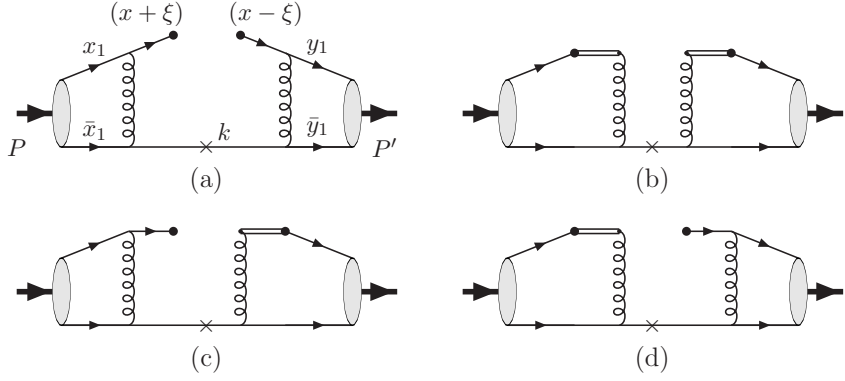}
\end{center}
\vskip -0.7cm \caption{\it Leading order contributions to the
generalized parton distribution $H_q(x,\xi,t)$ for pion at large
$x$. The crosses represent the intermediated states, and the
double lines for the eikonal term from the gauge link.}

\vskip -0.3cm
\end{figure}

The calculation of the diagrams in Fig.~1 is straightforward, and
the result is
\begin{eqnarray}
H_q^\pi(x,\xi,t)&=&\int\frac{d^2k_\perp}{(2\pi)^3} \frac{1}{(2
k\cdot P) (2 k\cdot P')}{\cal I}(x,\xi)\ . \label{hupi}
\end{eqnarray}
The integral ${\cal I}$ depends on the distribution amplitudes of
the initial and final states,
\begin{eqnarray}
{\cal I}(x,\xi)&=&(4\pi\alpha_sC_F)^2\int dx_1dy_1
\frac{\phi(x_1)\phi^*(y_1)}{\overline{x}_1\overline{y}_1}
\left(1+\frac{1}{\overline{x}_1-\frac{1-x}{1+\xi}}\right)
\left(1+\frac{1}{\overline{y}_1-\frac{1-x}{1-\xi}}\right) \ ,
\nonumber
\end{eqnarray}
where $C_F=4/3$ and $\overline{x}_1=1-x_1$. The integral becomes a
constant in the limit of $x\to 1$. In Eq.~(\ref{hupi}), the
denominator factors $2(k\cdot P)$ and $2(k\cdot P')$ in  come from
the gluon propagators in the diagrams. They depend on the momentum
transfer $t$ in general. However, if expanded at small $(1-x)$,
they become
\begin{eqnarray}
\frac{1}{2k\cdot P}&=&\frac{1-x}{\vec k_\perp^2
(1+\xi)}\left[1+\frac{(1-x)^2(1-\xi^2)t}{4(1+\xi)^2\vec
k_\perp^2}\right] \ ,\nonumber\\
\frac{1}{2k\cdot P'}&=&\frac{1-x}{\vec k_\perp^2
(1-\xi)}\left[1+\frac{(1-x)^2(1-\xi^2)t}{4(1-\xi)^2\vec
k_\perp^2}\right] \ , \label{prop}
\end{eqnarray}
which implys that there is no $t$-dependence in the leading order,
and any dependence will be suppressed by a factor of $(1-x)^2$.
Since the propagators are the only source of the $t$-dependence of
the GPD in Eq.~(\ref{hupi}), we conclude that at $x\to 1$ the GPD
$H(x,\xi,t)$ for pion has no $t$-dependence, and any
$t$-dependence must be suppressed by a factor of $(1-x)^2$. These
conclusions agree with the analysis in the impact parameter
dependent picture of the GPDs at $\xi=0$ \cite{Burkardt:2003ck},
while our results are valid for any finite value of $\xi$.

Collecting the above results, we have the GPD $H(x,\xi,t)$ for the
pion in the limit of $x\to 1$,
\begin{equation}
H_q^\pi(x,\xi,t)=\frac{(1-x)^2}{1-\xi^2}\int\frac{d^2k_\perp}{(2\pi)^3}
\frac{1}{(k_\perp^2+\Lambda^2)^2}{\cal I}\ . \label{pih}
\end{equation}
There is an infrared divergence where the transverse momentum
$k_\perp$ becomes soft. This divergence breaks the factorization
in principle. However, this does not change the power behavior.
Nevertheless, we include a regulator $\Lambda$ to regulate such
divergence \cite{Mueller:sg}. Like the forward parton
distributions, a power behavior is found here for the GPDs in
Eq.~(\ref{pih}). If we take $\xi=0$ and $t=0$, Eq.~(\ref{pih})
will reproduce the forward parton distribution of pion. So, in the
limit of $x\to 1$, the GPD $H(x,\xi,t)$ for pion can be related to
the forward parton distribution $q^\pi(x)$,
\begin{equation}
H_q^\pi(x,\xi,t)= \frac{1}{1-\xi^2} q^\pi(x)\ .
\end{equation}
We note that the above equality saturates the positivity
constraints \cite{Pobylitsa:2001nt,{Diehl:2003ny}} if we take the
power behavior for valence quark distribution $q^\pi(x)\sim
(1-x)^2$.

We turn now to the study of GPDs for the nucleon. Since the
leading Fock component of nucleon has three partons, many more
diagrams will contribute. Here we only show a particular diagram
in Fig.~2. There are two intermediate momenta, $k_1$ and $k_2$.
Similar to the above analysis for the pion case, we have a
hierarchy of scales: $\langle\vec k_\perp^2\rangle/(1-x)\gg
\langle\vec k_\perp^2\rangle \gg \Lambda^2_{\rm QCD}$ and
$\langle\vec k_\perp^2\rangle/(1-x)\gg (-t)$, where $\langle\vec
k_\perp^2\rangle$ represents the typical transverse momentum
scale, $\langle\vec k_\perp^2\rangle\sim \langle\vec
k_{1\perp}^2\rangle\sim \langle\vec k_{1\perp}^2\rangle$. Again,
we are only interested in the leading-order result, and neglect
any higher-order corrections in $(1-x)$.

The calculations are performed in the helicity bases for the
initial and final nucleon states $(\lambda',\lambda)$, in which
the following off-forward matrix elements are defined:
\begin{eqnarray}
{\cal H}_{\lambda'\lambda}&=&\!\!\frac{1}{2\sqrt{1-\xi^2}} \int
\frac{d\lambda}{2\pi} e^{i\lambda x} \left\langle
P',\lambda'\left|\overline{\psi}_q\left(-\frac{\lambda}{2}n\right)\!\!
\not\! n  {\cal L}
\psi_q\left(\frac{\lambda}{2}n\right)\right|P,\lambda\right\rangle
\nonumber\ . \label{proton}
\end{eqnarray}
The helicity non-flip amplitude has contributions from both $H$
and $E$ GPDs, while the helicity flip one only has the
contribution from $E$ GPD\cite{Diehl:2003ny},
\begin{eqnarray}
{\cal H}_{\uparrow\uparrow}&=&{\cal H}_{\downarrow\downarrow}=
H_q(x,\xi,t)-\frac{\xi^2}{1-\xi^2}E_q(x,\xi,t) \ ,\nonumber\\
{\cal H}_{\downarrow\uparrow}&=&-{\cal H}_{\uparrow\downarrow}^*=
\frac{\Delta^x+i\Delta^y}{2M_p(1-\xi^2)}E_q(x,\xi,t) \ .
\end{eqnarray}
We will show how the diagram in Fig.~2 contribute to these
amplitudes.

\begin{figure}[t]
\begin{center}
\includegraphics[height=2.0cm]{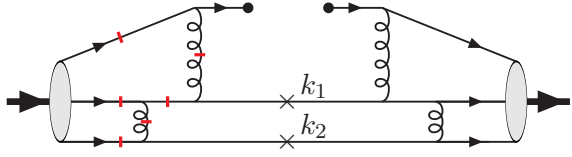}
\end{center}
\vskip -0.5cm \caption{\it The typical diagram contributing to the
nucleon GPDs at $x\to 1$. The quark helicity configuration is
$\{\uparrow\downarrow\uparrow\}$. The bars indicate the places
where we need to consider the internal transverse momentum
expansion to get helicity flip amplitude.} \vskip -0.3cm
\end{figure}

The helicity non-flip amplitude for the diagram of Fig.~2 has the
following form,
\begin{eqnarray}
{\cal H}_{\uparrow\uparrow}=\int\frac{d^2k_{1\perp}
d^2k_{2\perp}}{(2\pi)^3}\int\frac{d\alpha }{\alpha\beta
(1-x)}\left(\frac{1}{2P\cdot (k_1+k_2)}\frac{1}{2P'\cdot
(k_1+k_2)}\right)^2\frac{P'\cdot k_1}{P'\cdot k_2}{\cal
I}_p(x,\xi)\ , \label{e2}
\end{eqnarray}
where any other factors (such as color factors and coupling
constants, etc.) are included in the integral ${\cal I}_p(x,\xi)$.
This integral depends on the leading-twist distribution amplitudes
of the proton \cite{Braun:2000kw}, and will become a constant
integral at the limit of $x\to 1$. The longitudinal momentum
fractions of $k_1$ and $k_2$ are defined as $n\cdot
k_1=\alpha(1-x)$, $n\cdot k_2=\beta(1-x)$, and $\alpha+\beta=1$.
For the propagators, we make expansions at small $(1-x)$ as
before, for example,
\begin{equation}
\frac{1}{2P\cdot (k_1+k_2)}=\frac{\langle \vec
k_\perp^2\rangle}{1-x}(1+\xi)\left[1+{\cal
O}((1-x)^2)\frac{t}{\langle \vec k_\perp^2\rangle}+\cdots\right]\
,
\end{equation}
which again has no $t$-dependence at the leading order, and any
$t$ dependence is suppressed by a factor of $(1-x)^2$. All
propagators in Eq.~(\ref{e2}) have this property, and all other
diagrams which contribute to ${\cal H}_{\uparrow\uparrow}$ at the
leading order have the same dependence on $t$. The $t$ dependence
of nucleon GPDs is the same as that of the pion: there is no $t$
dependence and any dependence is suppressed by a factor of
$(1-x)^2$. In addition, every diagram contributes the same
dependence on $\xi$. Adding all of the contributions together, we
get
\begin{eqnarray}
{\cal
H}_{\uparrow\uparrow}=\frac{(1-x)^3}{(1-\xi^2)^2}\int\frac{d^2k_{1\perp}
d^2k_{2\perp}}{(2\pi)^3}\int\frac{d\alpha
}{\alpha\beta}F(\alpha,k_{1\perp},k_{2\perp},\Lambda){\cal I}_p\ ,
\end{eqnarray}
where the function $F$ is of order $1$ at $x\to 1$. Here we also
include a regulator $\Lambda$ in $F$ to regulate the infrared
divergences in the $k_{1\perp}$ and $k_{2\perp}$ integrations. If
we take $\xi=0$ and $t=0$, the above results will reproduce the
forward parton distribution at large $x$. That means we can have,
\begin{equation}
{\cal H}_{\uparrow\uparrow}=\frac{1}{(1-\xi^2)^2}q(x)\sim
\frac{(1-x)^3}{(1-\xi^2)^2}\ , \label{hq}
\end{equation}
at $x\to 1$.

Since hard scattering conserves the quark helicity, in order to
get the helicity flip amplitude ${\cal H}_{\uparrow\downarrow}$ we
must include non-zero orbital angular momentum either for the
initial or final states. In other words, we need to consider the
light-cone Fock components of hadrons with at least one unit of
orbital angular momentum \cite{Ji:2002xn}. The calculation for the
helicity flip amplitudes are much more complicated than that for
the helicity conserving ones. The method we are using follows
Ref.~\cite{Belitsky:2002kj} where the helicity flip Pauli form
factor was calculated in perturbative QCD. We will sketch the
method and summarize the main results, but skip the detailed
derivations.

First, we keep the internal transverse momenta $l_\perp$ of the
scattering partons in the hard partonic scattering amplitudes.
Then, we expand the amplitudes at small $l_\perp$. Since
$\Delta_\perp$ is the only relevant external transverse momentum,
the expansion of the amplitudes will be proportional to
$\vec\Delta_\perp\cdot \vec{l}_\perp$ or $\vec\Delta_\perp\times
\vec{l}_\perp$. Integrating these terms over $l_\perp$ with the
light-cone wave functions, we will get, e.g., $ \int d^2l_\perp
\vec\Delta_\perp\cdot \vec{l}_\perp (l_\perp^x+i
l_\perp^y)\psi^{(3,4)}\sim (\Delta_\perp^x+i\Delta_\perp^y)
\Phi^{(3,4)}$, and $\int d^2l_\perp \vec\Delta_\perp\times
\vec{l}_\perp (l_\perp^x+i l_\perp^y)\psi^{(3,4)}\sim
-i(\Delta_\perp^x+i\Delta_\perp^y) \Phi^{(3,4)}$, where
$\psi^{(3,4)}$ are the light-cone wave functions for the Fock
state with one unit orbital angular momentum \cite{Ji:2002xn}, and
$\Phi^{(3,4)}$ are the related twist-four distribution amplitudes
\cite{Braun:2000kw}. Thus, the final results of ${\cal
H}_{\uparrow\downarrow}$ depend on the twist-three and twist-four
distribution amplitudes of the nucleon.

We must consider the expansions for all propagators and quark wave
functions which have dependence on $l_\perp$. As an example, in
Fig.~2 we indicate all places where the $l_\perp$ expansion should
be considered if the initial state has one unit of orbital angular
momentum. These expansions will give additional power of
$(1-x)^2$, leading to the helicity flip amplitudes suppressed by
$(1-x)^2$. For instance, one of the gluon propagators in the
diagram of Fig.~2 has the following expansion,
\begin{equation}
\frac{1}{(k_2-x_3P-l_\perp)^2}=\frac{1}{(k_2-x_3P)^2}\left[1-\frac{\beta
(1-x)^2\vec \Delta_\perp\cdot
\vec{l}_\perp}{(1+\xi)^2\vec{k}_{2\perp}^2}\right]\ .
\end{equation}
Extracting the expansion coefficients, and combing with other
factors in the amplitude, we get the contribution to the helicity
flip amplitude ${\cal H}_{\downarrow\uparrow}$ from this term:
${\cal H}_{\downarrow\uparrow}\sim (1-x)^5/(1-\xi^2)^2(1+\xi)^2 $.
Adding the similar contribution from the final state expansion, we
get ${\cal H}_{\downarrow\uparrow}\sim
(1-x)^5/(1-\xi^2)^2(1/(1+\xi)^2+1/(1-\xi)^2)=
(1-x)^5(1+\xi^2)/(1-\xi^2)^4 $. All expansions result in the same
suppression of $(1-x)^2$. However, they do not contribute the same
dependence on $\xi$. For example, the quark wave function
expansions lead to ${\cal H}_{\downarrow\uparrow}\sim
(1-x)^5/(1-\xi^2)^4 $. In summary, the helicity flip amplitude
will have the following result at $x\to 1$,
\begin{equation}
{\cal
H}_{\downarrow\uparrow}\sim(\Delta_\perp^x+i\Delta_\perp^y)\frac{(1-x)^5}{(1-\xi^2)^4}f(\xi)\
.
\end{equation}
Here $f(\xi)$ represents an additional dependence on $\xi$, which
will depend on the input of the twist-three and twist-four
distribution amplitudes of the nucleon. From this, we deduce the
behavior of GPD $E_q(x,\xi,t)$ as,
\begin{equation}
E_q(x,\xi,t)\sim \frac{(1-x)^5}{(1-\xi^2)^3}f(\xi)\ .
\end{equation}
Comparing with Eq.~({\ref{hq}), we can neglect the $E_q$
contribution to the helicity non-flip amplitude, and then we have
$ H_q(x,\xi,t)\sim (1-x)^3/(1-\xi^2)^2$. So, in the limit of $x\to
1$, $H_q(x,\xi,t)$ can be related to the forward quark
distribution $q(x)$,
\begin{equation}
H_q(x,\xi,t)=\frac{1}{(1-\xi^2)^2}q(x) \ .
\end{equation}
Again this relation saturates the positivity constraint
\cite{{Diehl:2003ny},{Pobylitsa:2001nt}} for nucleon GPDs if the
forward quark distribution takes the power behavior at large $x$:
$q(x)\sim (1-x)^3$.

Before concluding, a few cautionary comments are in the order.
First, we omit the scale dependence of the GPDs. The scale
dependence at large $x$ is not just the simple DGLAP evolution
\cite{{Brodsky:1994kg},Manohar:2003vb}. In our calculations we
implicitly assume $Q^2(1-x)\gg\Lambda_{\rm QCD}^2$. Second, at the
limit of $x\to 1$ there exist $\alpha_s^n{\rm log}^m(1/(1-x))$ for
$m\le 2n$ series terms which need to be resummed, leading to a
Sudakov form factor suppression
\cite{{Lepage:1980fj},Mueller:sg,Manohar:2003vb,{Sterman:1986aj}}.
Third, the soft mechanism might contribute to the GPDs
\cite{Radyushkin:1998rt} at $x\to 1$. We did not include such
effects in our analysis.

In summary, we have studied generalized parton distributions at
$x\to 1$. We found that the pion's GPD $H_q^\pi(x,\xi,t)\sim
(1-x)^2/(1-\xi^2)$, and the nucleon's GPD $H_q(x,\xi,t)\sim
(1-x)^3/(1-\xi^2)^2$ and $E_q(x,\xi,t)\sim (1-x)^5/(1-\xi^2)^4
f(\xi)$. There is no $t$ dependence, and any dependence is
suppressed by a factor of $(1-x)^2$. These results can provide
important information on the GPDs' parameterizations.

We thank Xiangdong Ji for pointing out the $t$ dependence of the
GPDs at large $x$ and many other stimulating discussions. We also
thank Andrei Belitsky, Stan Brodsky, and Jianwei Qiu for useful
comments. This work was supported by the U. S. Department of
Energy via grants DE-FG02-93ER-40762.

\end{document}